\documentclass[prl,twocolumn,superscriptaddress,showpacs,citeautoscript,amsmath,amssymb,nobalancelastpage]{revtex4}
\usepackage[dvips]{graphicx}
\usepackage{dcolumn}
\usepackage{bm}
\usepackage{times}

\begin{document}

\title{Regular quantum plasmons in segments of graphene nanoribbons}

\author{Bao-Ji Wang}
\affiliation{Department of Physics, Tongji University, 1239 Siping Road,
Shanghai 200092, P. R. China}
\affiliation{College of Physics and Chemistry, Henan Polytechnic University,2001
Shiji Road, Jiaozuo 454000, P. R. China}

\author{San-Huang Ke} \email{shke@tongji.edu.cn}
\affiliation{Department of Physics, Tongji University, 1239 Siping Road,
Shanghai 200092, P. R. China}
\affiliation{Beijing Computational Science Research Center, 3 Heqing Road,
Beijing 100084, P. R. China}


\author{Hai-Qing Lin} 
\affiliation{Beijing Computational Science Research Center, 3 Heqing Road,
Beijing 100084, P. R. China}

\begin{abstract}
Graphene plasmons have advantages over noble metal plasmons, such as
high tunability and low loss. However, for graphene nanostructures 
smaller than 10\,nm, little is known about their plasmons or
whether a regular plasmonic behavior exists, despite their potential 
applications. Here, we present first-principles calculations 
of plasmon excitations in zigzag graphene nanoribbon segments.
Regular plasmonic behavior is found: Only one plasmon
mode exists in the low-energy regime ($<$1.5\,eV). 
The classical electrostatic scaling law still approximately holds
when the width ($W$) is larger than $\sim$1.5\,nm but totally fails when 
$W$$<$1.5\,nm due to quantum effects. The scaling with different doping densities
shows that the plasmon is nearly free-electron plasmon instead of Dirac plasmon.  
\end{abstract}
\maketitle

Collective electron oscillations in nanostructures form the localized surface
plasmon resonance upon interaction with light. This resonance can confine
electromagnetic energy down to deep sub-wavelength regions (truly nano-meter
scales) and, consequently, enhance the intensity of an incident light wave by
several orders of magnitude. Therefore, it is finding promising applications of a wide
range, such as nonlinear optics \cite{Danckwerts2007-026104,Ozbay2006-311},
single-molecule sensing \cite{Talley2005-1569}, and optical harvesting of
nanometer-sized objects \cite{Juan2011-349}. 
 
For traditional noble metal plasmon, its frequency is difficult to tune since
the electron density cannot be altered. Additionally, the plasmon scattering
time in metals is usually very short (typically tens of femtoseconds) and thus the
plasmons decay quickly. These drawbacks limit the performance of metamaterials
and transformation optical devices \cite{Boltasseva2011-290} and greatly
motivate people to explore plasmons in a newly available material with unique
properties \cite{Geim2007-183}: graphene. 
Its unique electronic properties are mainly due to the very peculiar
band structure, with the $\pi$ and $\pi^*$ bands showing linear dispersion around
the Fermi level, where they touch with each other at a single
point $K$ in the Brillouin zone \cite{Geim091530,Zhang05201}. Graphene can
be driven away from the charge neutrality point (CNP) and tuned into a metallic regime
by applying a voltage externally \cite{Novoselov2004-666} or electric gating
\cite{Wang2008-206}.
Fermi energies of the order of $\sim$1\,eV from the CNP are
currently attainable \cite{Efetov2010-256805}, which corresponds to electron doping densities up to
10$^{14}$cm$^{-2}$.
Compared to noble-metal plasmons, graphene plasmon possesses outstanding
advantages, such as highly tunable frequency, reduced loss, and also much
larger oscillator strength compared to plasmons of semiconductor
2-dimensional electron gas (2DEG) \cite{Jablan2009-245435,Ju2011-630}.

Recently, plasmon resonances in graphene have been probed using
inelastic electron scattering spectroscopy \cite{Liu2008-201403,Tegenkamp2011-012001}
and inelastic scanning tunnelling microscopy \cite{Brar2010-036805}.
The electrostatic tunability of graphene plasmons has been demonstrated by 
the terahertz light absorption in engineered microribbon arrays
\cite{Ju2011-630} and the infrared scattering strength of a tip situated 
near graphene \cite{Fei2011-4701}. All of these pioneering work mainly focused on
systems with biggish dimensions, whose plasmon frequencies are believed to obey
the electrostatic scaling law (ESL) \cite{Ju2011-630,Thongrattanasiri2012-1766}: 
$\omega_{p}\propto{W}^{\text{-}1/2} \times n^{1/4}$
with $W$ and $n$ being the width and carrier density of the
system, respectively. This scaling law stems from the nature of 2DEG of the
massless Dirac electrons. 
Besides the graphene microstructures and arrays of big sizes, very recently some
small graphene nanostructures, like graphene nanoisland, have also been synthesized \cite{Phark118162}. 
These small graphene clusters may be chemically doped and could have promising
applications, for example, for achieving optical sensing of chemical changes in a fluid surrounding the
cluster. To have a high sensitivity, these graphene nanoclusters should be made
as small as possible so that the adding or removing of one single electron will
cause a large change in the carrier density and inducing observable plasmon shifts. 

Unlike graphene nanostructures larger than $\sim$10\,nm where the
plasmonic behavior may be still described by the classical electrodynamics
\cite{Thongrattanasiri2012-1766}, 
graphene nanoclusters smaller than $\sim$10\,nm may have
quite different behavior due to the quantum confinement effects and
their discrete energy levels instead of continuous energy bands present in
periodic structures. As a result, the classical ESL relation
may fail. Therefore, it is interesting to explore the new phenomena
in the quantum regime and the physics underlying. 
So far, some computational studies \cite{Thongrattanasiri2012-1766} based on the $\pi$-only tight-binding modeling
have been reported for circular graphene nanodisks with diameters from 2 to
24\,nm. It was found that when the diameter of a nearly
circular disk is smaller than $\sim$10\,nm, its plasmonic behavior is very irregular, showing many
scattered resonance peaks in the energy window of 0 -- 0.6 eV
probably due to the complicated irregular edge geometry. 
This largely limits its potential applications. 
Therefore, finding small graphene nanostructures with regular plasmonic
behavior is desirable.

Another interesting question concerning graphene nanostructures is the nature of
their plasmons. For graphene microribbons,
the plasmon is believed to be Dirac plasmon due to the linear dispersion \cite{Ju2011-630}.  
Down to the nanometer scales as in the case of graphene nanoribbons, this linear
dispersion will not hold. Additionally, electron doping may not only shift the
position of the Fermi level but also change the electronic structure.
Recent {\it ab initio} band-structure
calculations for graphene and graphane nanoribbons showed 
that a heavy doping may lower the energy of the nearly
free-electron states to the vicinity of the Fermi level \cite{Liu1122,Hu2010}.
All these factors will affect the plasmonic behavior of graphene
nanostructures and the nature of their plasmons  
is still an open question to answer.   

In this work, we present, for the first time to our best knowledge, 
a full first-principles study of plasmon
excitations in doped zigzag graphene nanoribbon segments (ZGNRSs) of
different widths and doping densities
by using the time-dependent density functional theory (TDDFT) calculation 
which have been extensively used for studying electronic excitations
in nanostructures \cite{Wang2012,Lehtovaara2011-154104, Marinopoulos2002-076402, 
Marinopoulos2003-046402, Tachikawa2009-2101, Gao2011-1009, Yan2008-235413}. 
Our calculation shows that the plasmonic behavior of ZGNRSs is very regular:
The photoabsorption spectra show only three peaks in the energy window of 0 -- 1.5 eV
and only the lowest one is due to the plasmon excitation, as revealed directly
by the frequency-resolved induced charge density (FRID). 
The frequency of the only plasmon mode ($\omega_p$) as a function of the doping
density ($n$) and ribbon width ($W$)
is investigated. The variation with $W$ tends to approach the classical ESL
\cite{Hwang2007-205418,Ju2011-630} when $W>1.5$\,nm but shows a totally
different behavior when $W<1.5$\,nm because of a dominant quantum effect.
The variation with $n$ shows that the plasmon in ZGNRSs is 
nearly free-electron plasmon instead of Dirac plasmon.

\begin{figure}[tb]
\includegraphics[width=8.0cm,clip]{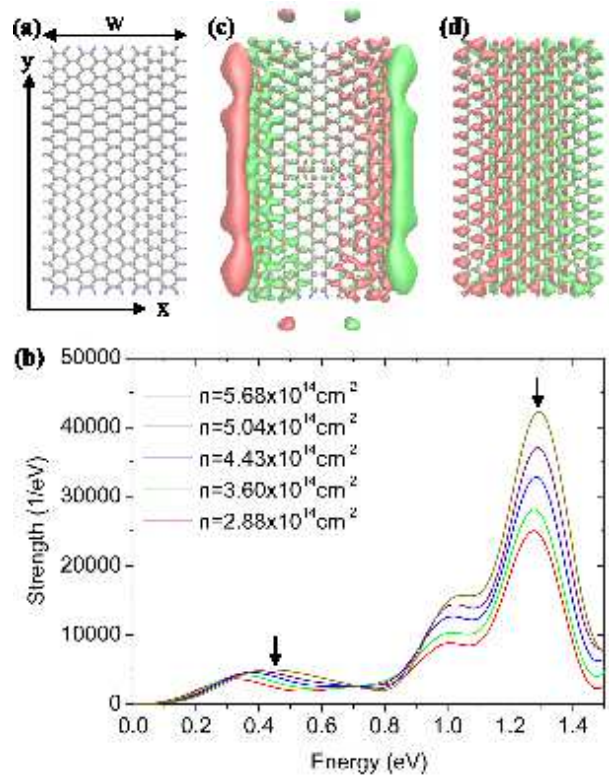}
\caption{\label{fig:10-ZGNR-as} (a) The structure and (b) the dipole response of the 10-ZGNRS
with different doping density $n$ as indicated. (c) and (d) are the FRID of the
left and right resonance peaks in (b), respectively, with red and green colors
denoting the positive and negative FRID, respectively. Note that the two FRID
distributions are very different from each other, indicating different natures of the two
resonances.}
\end{figure}

The computational details of our work are as follows.
We study ZGNRSs of different widths ($W$) which can be also denoted by the
number of the zigzag carbon chains ($m$) contained (labeled by $m$-ZGNRS).
In this work, $m$=2, 4, 6, 8, and 10 with corresponding width being 5.12, 9.38, 13.64, 17.90, and 22.16{\AA},
respectively, are considered.
The length-width ratio is chosen to be about 2 (see Fig.\ref{fig:10-ZGNR-as}(a) 
for the structure of the 10-ZGNRS; Different length-width ratios are also checked 
and no qualitative effect is found). The edges of the ZGNRSs are saturated by H atoms.
The ZGNRSs are placed in the $xy$-plane with zigzag chains being in the 
$y$-direction (see Fig.~\ref{fig:10-ZGNR-as}(a)).              
A perturbation caused by an impulse field is applied along $x$-axis to activate
dipole oscillations.
Different doping densities of the order of 10$^{14}$\,cm$^{-2}$ are 
considered (see Fig.~\ref{fig:10-ZGNR-as}(b)) since undoped ZGNRSs do not show
visible plasmon resonances. We note that the minimum doping density required to obtain
a visible plasmon oscillation decreases with the increasing $W$. 

Our calculations are carried out with the TDDFT implemented in the real space
and real time \cite{Marques2003-60}.
The carbon and hydrogen atoms are described by the Troullier-Martins
pseudopotentials \cite{Troullier1991-1993} and the local density approximation
(LDA) and the adiabatic LDA are adopted to describe the electron exchange and
correlation for the ground-state and excited-state calculations, respectively. 
This scheme has been extensively tested and used  
in predicting the photoabsorption and plasmon resonances of clusters 
\cite{Vasiliev1999-1919,Onida2002-601,Marinopoulos2002-076402,Yan2008-235413,Wang2012}.
Technically, a grid in real space which is defined by assigning a sphere around
each atom with a radius of 6{\AA} and a uniform mesh grid of 0.3{\AA} is adopted to
describe the wavefunction and charge density. To obtain the excitation spectrum,
the system is impulsed from its initial ground state with a very short
delta-function-like perturbation, and then the time-dependent Kohn-Sham equation
is evolved in the real space and real time for a certain period of time.
Specifically, an electronic wave packet is evolved for 10,000 time steps with
each being 0.003 $\hbar$/eV long. After the real-time propagation, the
photoabsorption spectrum is extracted by Fourier transforming the time-dependent
dipole strength. Furthermore, a 3D image of the FRID distribution is obtained 
for each resonance in the spectrum. This is
achieved by Fourier transforming the time series of the total induced charge
density at each resonance frequency for every real-space mesh grid. 

Fig.~\ref{fig:10-ZGNR-as}(b) shows photoabsorption spectra for the
10-ZGNRS with different electron doping density ($n$). The first thing to note is that there are only
three peaks in the energy window 0 -- 1.5\,eV. Along with the increase of $n$,
the frequency of the two high-energy peaks 
do not change while the low-energy peak bluesifts remarkably. 
The different behavior with the varying $n$ reflects different natures
for the two kinds of resonance. As described by the ESL, the plasmon frequency of a 2DEG should
bluesift as $n$ increases, indicating that the low-energy
resonance may be due to plasmon excitation while the high-energy resonances are
not. To check this, we draw the FRID for the low-energy and the dominant
high-energy peaks in Figs.~\ref{fig:10-ZGNR-as}(c) and (d), respectively. 
One can see that the FRID of the low-energy peak shows
indeed a collective charge oscillation with the induced density localized around
the surfaces (i.e., the two edges) -- a plasmon resonance.
However, the FRID of the high-energy peak shows
an positive and negative induced density distribution alternatively
through out the whole ZGNRS plane, indicating that it is due to 
the local dipole oscillation associated with the
$\sigma$-states. Our calculations for other ZGNRSs of different widths show
qualitatively the same result. Thus, we find that in ZGNRSs there exists only one plasmon
mode in low-energy regime. This is in striking contrast with the
case of circular graphene nanodisk \cite{Thongrattanasiri2012-1766} where many
scattered plasmon modes are found in the energy window 0 -- 0.6\,eV. 
In the following, we focus only on this plasmon mode and how it behaves with the
varying $W$ and $n$. 

\begin{figure}[tb]
\includegraphics[width=8.0cm,clip]{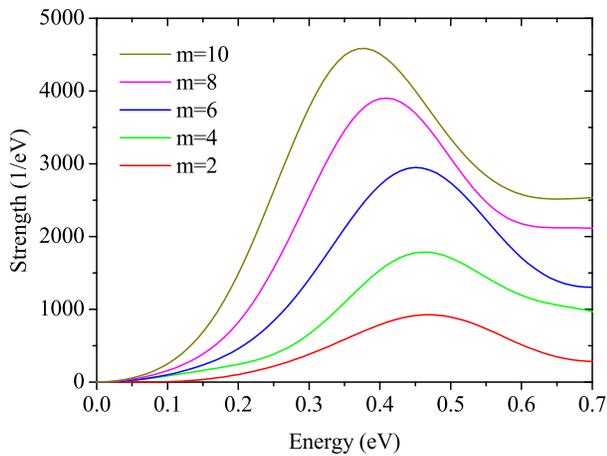}
\caption{\label{fig:n-ZGNR-as} The dipole response of the doped ZGNRS
of different widths ($m$=2, 4, 6, 8, 10) for $n$=4.43$\times$10$^{14}$ cm$^{-2}$. 
Note the redshift of the plasmon resonance with the increasing width.
}
\end{figure}

\begin{figure}[tb]
\includegraphics[width=8.0cm,clip]{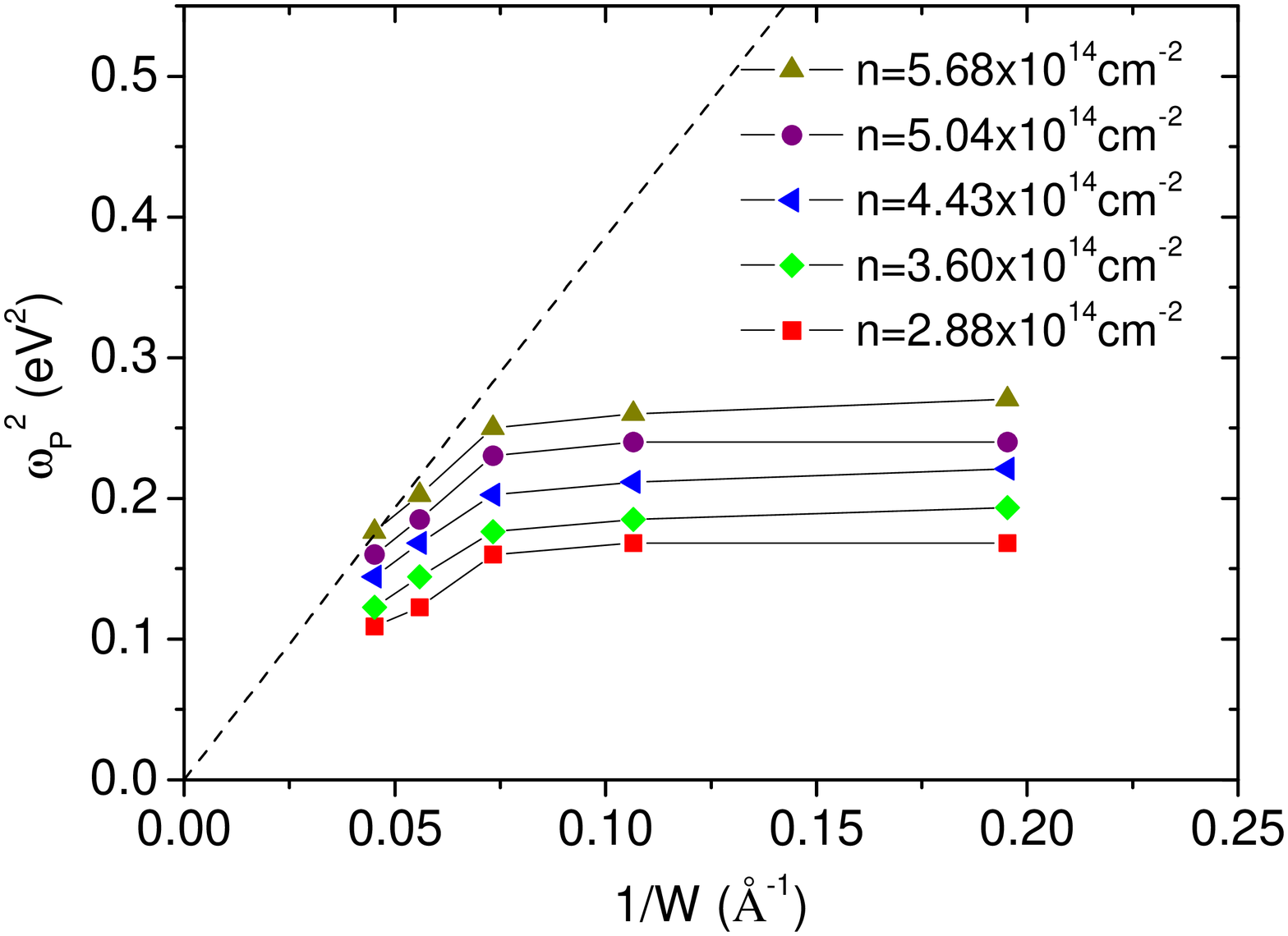}
\caption{\label{fig:wp-w} Variation of $\omega_{p}^2$ with $W^{-1}$
for different doping densities as indicated.
As a comparison, the simple proportional relation $\omega_{p}^{2} \propto
W^{-1}$ of the ESL is also plotted with a dashed line through the data
point of the maximum $n$ and $W$.
Note that for large $W$ the variation of $\omega_{p}^2$ is close to the
ESL while for small $W$ $\omega_{p}^2$ becomes almost a
constant.
}
\end{figure}

\begin{figure}[tb]
\includegraphics[width=8.0cm,clip]{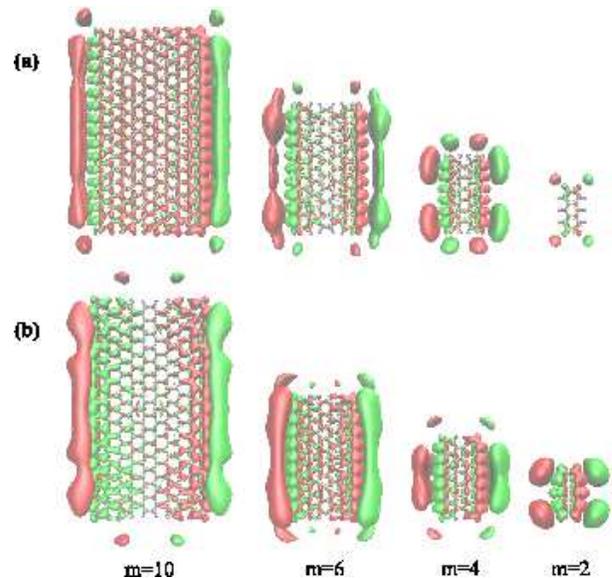}
\caption{\label{fig:FRID-w} The FRID distribution of the ZGNRSs with different
widths as indicated, for two doping densities: (a) $n$=2.88$\times$10$^{14}$ cm$^{-2}$ and (b)
$n$=5.68$\times$10$^{14}$ cm$^{-2}$. Note that the FRID is gradually driven to
the end of the ZGNRS as $W$ decreases.  
}
\end{figure}

Let us start with the plasmon frequency ($\omega_{p}$) as a function of $W$. 
In Fig.\ref{fig:n-ZGNR-as} we plot the photoabsorption spectra of the $m$-ZGNRS
($m$=2, 4, 6, 8, and 10) for the doping density
$n$=4.43$\times$10$^{14}$ cm$^{-2}$. One can see that the plasmon frequence
redshifts with the increasing $W$ while the dipole strength increases.
The redshift in frequency can be understood by considering the decreased energy gap involved in the
plasmon excitation and the enhancement of strength results from
more electrons participating in the dipole oscillation. Similar behavior was
also found in other 2D systems, like 2D sodium atomic planes \cite{Wang2012}, 
and is consistent qualitatively with the ESL of 2DEGs ($\omega_{p}^{2} \propto 1/W$).  

To show quantitatively the relation between $\omega_{p}$ and $W$, we plot in
Fig.\ref{fig:wp-w} the variation of $\omega_{p}^2$ with $W^{-1}$ for different
doping densities. The simple proportional relation of the ESL is also drawn
with a dashed line through the data point of the maximum $n$ and $W$ for a
comparison.
It appears that when the ZGNRS is wide ($m \ge 6$) its $\omega_{p}^2$ as a
function of $W^{-1}$ tends to approach the ESL and this agreement holds for
all the different doping densities considered. This indicates that the ESL
is still applicable to some extent even for ZGNRS as narrow as $\sim$2nm. 
However, when the ZGNRS becomes very narrow ($m < 6$) the variation of
$\omega_{p}^2$ deviates significantly from the ESL: $\omega_{p}^2$ is now
little affected by $W$. This shows that for very narrow ZGNRS the quantum finite-size
effect plays an important role and the classical ESL is not applicable anymore.
To understand this deviation from the ESL, 
we plot in Fig.\ref{fig:FRID-w} the FRID for four different widths ($m$=2, 4, 6,
10) and two different doping densities. For large width ($m$=10) 
the FRID is mainly distributed around the two long sides of the ZGNRS,
showing a (bulk) central plasmon mode which should be approximately described by
the ESL. However, when the width is reduced to $m$=2 the ZGNRS
will become a quasi-1D system and the FRID is now driven to the two ends, showing an end
plasmon mode. As previously demonstrated in the case of atomic chains
\cite{Yan2008-235413,Wang2012}, this end mode 
is a pure quantum effect and cannot be described by the ESL: Its frequency 
is almost a constant. This spatial evolvement of the FRID, 
from the central mode to the end mode, reveals that
emerging quantum effect takes control gradually as the width is reduced. 

It should be noted that the emergence of the end mode is completely due to the
quantum finite-size effect and is not affected by the doping density. This is
evident in Fig.\ref{fig:FRID-w}: 
The large increase of $n$ from Fig.\ref{fig:FRID-w}(a) to
Fig.\ref{fig:FRID-w}(b) merely enhances the central or the end
mode but affects little the nature of them.

\begin{figure}[tb]
\includegraphics[width=8.0cm,clip]{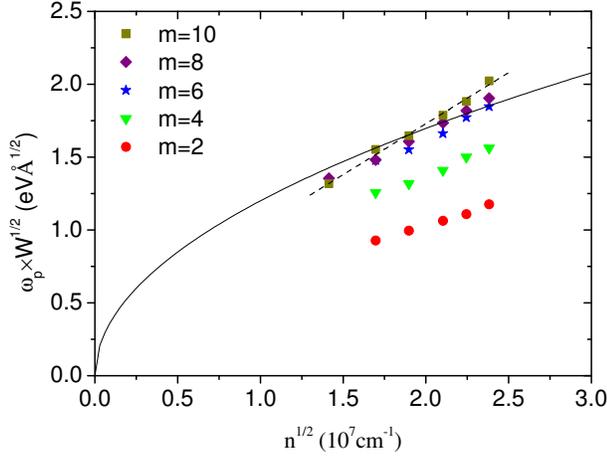}
\caption{\label{fig:normalized-1} Plasmon frequency $\omega_{p}$ normalized by
$W^{-1/2}$ as a function of $n^{1/2}$ for ZGNRSs of different
widths (data points), in comparison with the ESL scaling of Eq.(1) for $m$=10 (solid line).
Note the different scaling behavior when the ZGNRS is very narrow ($m$=2, 4). The
dashed line shows the trend for the large width ($m$=10).
}
\end{figure}

\begin{figure}[tb]
\includegraphics[width=8.0cm,clip]{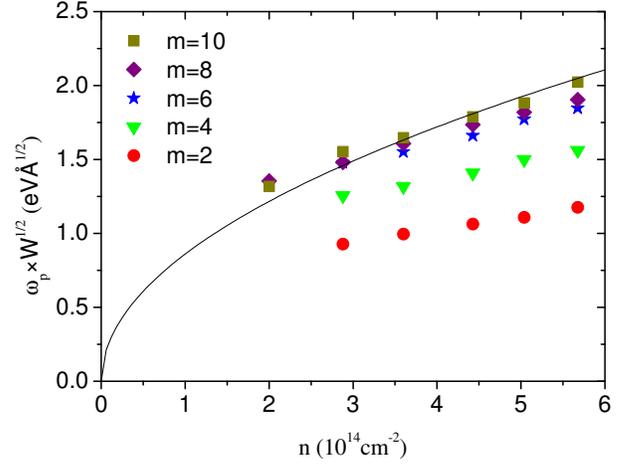}
\caption{\label{fig:normalized-2} Plasmon frequency $\omega_{p}$ normalized by 
$W^{-1/2}$ as a function of $n$ for ZGNRSs of different widths (data points), in
comparison with the ESL scaling of Eq.(2) for $m$=10 (solid line). 
Note the good agreement between the solid line and the data points for large $m$ and
$n$.
}
\end{figure}

Next, we investigate the plasmon frequency as a function of the doping density.
According to the ESL, the frequency of 2D plasmon will scale as
\begin{equation}
\omega_{p} \times {W}^{1/2} \propto n^{1/4}
\end{equation}
for massless-electron (Dirac) plasmon or
\begin{equation} 
\omega_{p} \times {W}^{1/2} \propto n^{1/2}
\end{equation}
for free-electron plasmon. 
To date, (doped) graphene plasmon is thought to be Dirac plasmon
and the experimental data for graphene microribbons reported in
Ref.\onlinecite{Ju2011-630} were claimed to fit Eq.(1) very well. 
It is interesting to see to what extent
this Dirac ESL scaling still holds for the much smaller ZGNRSs.
To have a detailed comparison, we plot $\omega_{p}\times W^{1/2}$ as a function
of $n^{1/2}$ in Fig.\ref{fig:normalized-1} and compare the data points to the
fitted parabola (only for $m$=10), as did in the same way in Fig.3(b) of 
Ref.\onlinecite{Ju2011-630}. The first thing to note is that the normalized
scaling behavior strongly depends on the width. When the width is
very small ($m$=2, 4) the scaling is very different for different widths while for large
widths ($m$=6, 8, 10) the scaling behavior tends to converge. This is a result 
of the quantum finite-size effect discussed previously.
If one compare carefully the data points to the fitted parabola (for $m$=10) in
Fig.\ref{fig:normalized-1} one can see that the trend of the data points (indicated
by a dashed line) is not well consistent with the parabola: The slope of the
dashed line is remarkably larger than that of the parabola. This discrepancy
from the scaling relation of Eq.(1) indicates that the ZGNRS plasmon may not be Dirac plasmon.
Interestingly, a somewhat similar discrepancy also exists in the experimental
result for graphene microribbons (see Fig.3(b) of Ref.\onlinecite{Ju2011-630}), which may 
need further experiment to clarify. 

\begin{figure}[tb]
\includegraphics[width=8.0cm,clip]{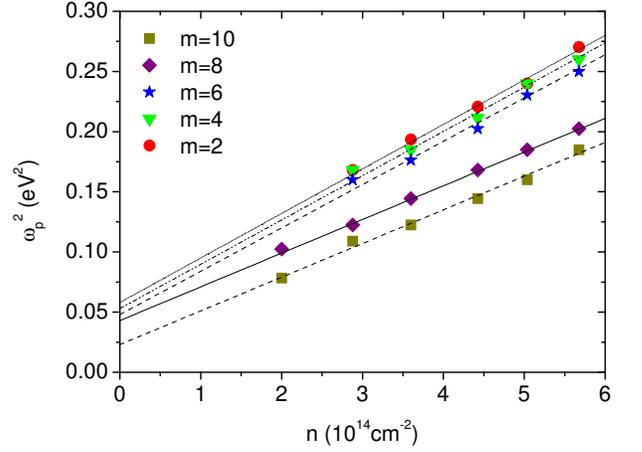}
\caption{\label{fig:wp^2-n} Variation of $\omega_{p}^2$ with doping density
$n$ for ZGNRSs of different widths. The straight lines show the fitted tend.
}
\end{figure}

To check the scaling relation of Eq.(2), we plot $\omega_{p}\times W^{1/2}$ as a function
of $n$ in Fig.\ref{fig:normalized-2} and compare the data points to the
fitted parabola (only for $m$=10). 
It can be seen that now the data points are in better agreement with the 
parabola, especially for the higher doping densities. This shows that Eq.(2) 
describes more reasonably the scaling behavior. To see this more directly, we plot
$\omega_{p}^2$ as a function of $n$ for different
widths in Fig.\ref{fig:wp^2-n}. It is evident that all the data points are in very good
agreement with the fitted straight lines, indicating that Eq.(2) holds very well 
for plasmons in ZGNRSs. This result suggests that the ZGNRS plasmon is nearly
free-electron plasmon instead of Dirac plasmon.   

In summary, by using {\it ab initio} time-dependent density functional theory calculation in the time
domain, we have investigated the collective electronic excitations in doped
segments of zigzag graphene nanoribbons with different widths and doping
densities. It has been found that the plasmonic behavior is very regular in the
low-energy regime: Only one plasmon mode exists in the energy window of
0--1.5\,eV. The variation of its frequency with the width tends to approach the
classical electrostatic scaling law for widths larger than $\sim$1.5\,nm but shows a totally
different behavior for widths smaller than $\sim$1.5\,nm due to a dominant quantum effect.
The variation of the plasmon frequency with the doping density shows that the
plasmon is nearly free-electron plasmon instead of Dirac plasmon.

We would like to thank Jin-Long Yang at USTC for helpful discussion. 
This work was supported by the National Natural Science Foundation of China 
under Grant No. 11174220 and by the MOST 973 Project under Grant No.
2011CB922204, and by King Abdulaziz University (KAU) under Grant No.
(49-3-1432/HiCi), as well as by the Shanghai Pujiang Program under Grant No. 10PJ1410000.


\end{document}